\documentstyle[pra,epsf,aps,twocolumn]{revtex}
\def\qn{x}
\def\phin{q}
\def\lfrac#1#2{#1/#2}
\def\db{\,\, {\bar{} \!\!d}\!\,\hspace{0,5pt}}
\def\comment#1{}
\def\lfrac#1#2{#1/#2}

\begin{document}
\sloppy
\title{Rules for Integrals over Products of
Distributions from
Coordinate Independence of Path Integrals}
\author{H.~Kleinert\thanks{E-mail: kleinert@physik.fu-berlin.de} and
     A.~Chervyakov\thanks{On leave from LCTA, JINR, Dubna, Russia
                   E-mail: chervyak@physik.fu-berlin.de}
                      \\ Freie Universit\"at Berlin\\
          Institut f\"ur Theoretische Physik\\
          Arnimallee 14, D-14195 Berlin}
\maketitle
\begin{abstract}
In perturbative calculations
of quantum mechanical path integrals in
curvilinear coordinates,
one encounters Feynman
diagrams involving
multiple temporal integrals over products of distributions,
which are
mathematically undefined.
In addition, there are
terms proportional to powers $ \delta(0) $
from the measure of path integration.
We derive simple rules
for their evaluation
from the natural requirement
of
coordinate independence of
the path integrals.
\end{abstract}
\section{Introduction}
While quantum mechanical path integrals in curvilinear coordinates
can be defined  uniquely
and independently of the choice of coordinates
within the time-sliced formalism
\cite{1}, a perturbative definition
on a continuous time axis
poses problems.
To exhibit the difficulties, consider the associated partition function
calculated for periodic paths on the imaginary-time axis $\tau $:
\begin{equation}
Z = \int  {\cal D} \phin  (\tau) \sqrt{g} \,
e^{-{\cal A} [\phin ]},
\label{@1}\end{equation}
where ${\cal A} [\phin ]$ is the euclidean action
with the general form
\begin{equation}
{\cal A} [\phin ]=\int d\tau \left[ \frac{1}{2}g_{\mu \nu }(q(\tau ))
\dot q^\mu(\tau )
\dot q^\nu(\tau )+V(q(\tau ))\right].
\label{@}\end{equation}
The dots denote $\tau $-derivatives, $g_{\mu \nu }(q)$ is
a metric, and $g=\det g$
its determinant.
The path integral may formally
be defined perturbatively as follows:
The metric
$g_{\mu \nu }(q)$ is expanded
around some point $q_0^\mu$
in powers of $ \delta q^\mu\equiv q
^\mu-q_0^\mu$.
The same thing is done with the potential $V(q)$.
After this, the action ${\cal A} [\phin ]$
is separated into a free  part
${\cal A}_0[q_0; \delta \phin ]\equiv
\frac{1}{2}g_{\mu \nu }(q_0))
\dot q^\mu\dot q^\nu+\frac{1}{2} \omega ^2 \delta q^\mu \delta q^ \nu$,
and an interacting part
${\cal A}_{\rm int}[q_0; \delta \phin ]\equiv
{\cal A}[ \phin ]-
{\cal A}_0[q_0; \delta \phin ]$.

A first problem is encountered
in the square root in the functional integration measure in
(\ref{@1}).  Taking it into the exponent and expanding it
in powers of $ \delta q$ we define
an effective action$ {\cal A}_{ \sqrt{g} }=-\frac{1}{2} \delta (0)
\log [ g(q_0+ \delta q)/g(q_0)]$,
which contains the
$ \delta $-function at the origin
$ \delta (0)$.
It represents formally
the inverse infinitesimal lattice spacing on the time axis, and is
equal to the infinite number
$\delta (0)\equiv \int d p/(2\pi)$.

The second problem arises in the
expansion of $Z$ in powers of the interaction,
Performing all Wick contractions,
$Z$ is expressed as
a sum of loop diagrams.
There are interaction terms involving $\dot q^2 q^n$
which lead to Feynman integrals
over products of distributions.
The diagrams contain
three types of lines representing
the correlation functions
\begin{eqnarray}
\Delta (\tau -\tau ')&\equiv& \langle \phin (\tau )\phin (\tau ')\rangle=
\hspace{0mm}\raisebox{-1mm}{\mbox{\input 1.tps }} ,~\label{p1}\\
\partial _\tau \Delta(\tau -\tau ')&\equiv &\langle \dot \phin (\tau )\phin (\tau ')\rangle
=\hspace{0mm}\raisebox{-1mm}{\mbox{\input 3.tps }} ,~\label{p2}\\
 \partial_ \tau \partial_{\tau '}\Delta (\tau-\tau ')&\equiv&
\langle \dot \phin (\tau ) \dot\phin (\tau ')\rangle
=\hspace{0mm}\raisebox{-1mm}{\mbox{\input 2.tps }}.~\label{p3}
\label{@}\end{eqnarray}
The right-hand sides define the line symbols
to be used in Feynman diagrams for the interaction terms.

Explictly,
the first correlation function reads
\begin{equation}
 \Delta (\tau ,\tau ')=\frac{1}{2\omega}e^{-\omega|\tau -\tau '|}.
\label{p4}\end{equation}
The second correlation function (\ref{p2})
has a discontinuity
\begin{equation}
\partial _\tau \Delta(\tau ,\tau ') =
     - \frac{1}{2} \epsilon (\tau - \tau ') e^{-\omega|\tau -\tau '|} ,
\label{p5}\end{equation}
where
\begin{equation}
\epsilon (\tau - \tau ')\equiv-1+ 2\int_{-\infty}^\tau  d\tau''  \delta (\tau'' -\tau ')
\label{p6}\end{equation}
is a distribution
which vanishes at the origin and is equal to $\pm1$ for
positive and negative arguments, respectively.
The third correlation function (\ref{p3}) contains a
$ \delta $-function:
\begin{equation}
 \partial_ \tau \partial_{\tau '}\Delta (\tau, \tau ') =
  \delta(\tau -\tau ') - \frac{\omega}{2}e^{-\omega|\tau -\tau '|} ,
\label{p7}\end{equation}
The temporal
integrals  over products of such distributions
are undefined \cite{qft}.

In this note we  define them uniquely
by setting  up
rules between these and integrals over products of
nonsingular correlation  functions
$ \Delta (\tau - \tau')$, plus integrals
over pure products of $ \delta $-functions.
These will be defined uniquely
by the requirement of
coordinate invariance of the path integral (\ref{@1}).

The internal consistency of these definitions
is ensured by previous work
of the present authors.
In Ref.~\cite{2},
we have shown that Feynman integrals in {\em momentum
space\/} can be uniquely defined
as $ \epsilon \rightarrow 0$ -limits of $1- \epsilon $-dimensional
integrals via an  analytic  continuation
\`a la
't~Hooft and M.~Veltman \cite{4}.
This definition makes
path integrals coordinate independent.
In Ref.~\cite{3} we have given rules
for calculating the same results directly
from the Feynman integrals in the $1- \epsilon $ -dimensional time space.

The present approach has the important advantage
making superfluous the somewhat tedious
analytic continuation to $1- \epsilon $  dimensions.
In fact, it does not require specifying any regularization scheme.
In addition, it gives a foundation of a
new and general mathematics
of extending the theory of distributions
from a linear space to products.

\section{Model System}
The announced derivation of the
identities will be based on the
requirement
of  coordinate independence
of the
exactly solvable
path integral
of a point particle of unit mass in a harmonic
potential $\omega^2 \qn ^2/2$,
 over a large imaginary-time interval $ \beta $,
\begin{equation}
  Z_ \omega  = \int  {\cal D} \qn  (\tau)\,
e^{-{\cal A}_{ \omega } [\qn ]}
= e^{- \rm Tr \log (-\partial^2 + \omega^2)} =
  e^{-\beta\lfrac{\omega}{2}}.
\label{mq2}\end{equation}
The action is
\begin{equation}
{\cal A}_ \omega   = \frac{1}{2}\,\int\, d \tau \left[\dot \qn ^2(\tau ) + \omega^2 \qn ^2(\tau )
\right].
\label{m1}\end{equation}
 A coordinate transformation
turns  (\ref{mq2}) into a path integral of the type (\ref{@1})
with a
singular perturbation expansion.
From our work in Refs.~ \cite{2,3}
we know
that all
terms in this expansion vanish in dimensional regularization.
Here we shall {\em require\/} the vanishing
to find the desired identities for integrals over products of distributions.

For simplicity we assume the coordinate transformation
to preserve the symmetry $\qn \leftrightarrow -\qn $ of the initial oscillator,
such its power series expansion starts out like
$\qn  (\tau)=f(\phin (\tau)) = \phin  - {g}\phin ^3/3 + {g^2}a\phin ^5/5 - \cdots~$,
where $g$ is a smallness parameter, and $a$ an extra parameter.
We shall see that the identities are independent of $a$, such that $a$
will merely
serve
to check the calculations.
The transformation changes the partition function
(\ref{mq2})
into
\begin{equation}
  Z = \int  {\cal D} \phin  (\tau)\,
e^{-{\cal A}_{J} [\phin ]}
e^{-{\cal A} [\phin ]},
\label{m2}\end{equation}
where is $ {\cal A}_{} [\phin ]$ is the transformed action,
whereas
 $ {\cal A}_J[\phin ]$
 an effective action coming from the
Jacobian of the coordinate transformation:
\begin{equation}
 {\cal A}_J[\phin ]=
-\delta (0)\int d \tau\,\log \,\frac{\delta f(\phin (\tau))}{\delta \phin (\tau)}.
\label{m4}\end{equation}

The transformed action is decomposed into a free
part
\begin{equation}
    {\cal A}_{ \omega } [\phin ]= \frac{1}{2}\,\int\, d \tau [\dot \phin ^2(\tau )
+\omega^2 \phin ^2 (\tau)] ,
\label{m6}\end{equation}
and an interacting part, which reads to
 second order in  $g$:
\begin{eqnarray}
&&\!\!~
{\cal A}_{\rm int} [\phin ] = \frac{1}{2} \int d \tau
\left\{-g\left[ 2\dot \phin ^2  (\tau) \phin ^2 (\tau)
+ \frac{2\omega^2}{3}\phin ^4 (\tau) \right]\right.
\nonumber\\&&\left.  ~
 + g^2
\left[\left(1 + 2a\right) \dot \phin ^2 (\tau)  \phin ^4 (\tau)
+ \omega^2 \left(\frac{1}{9} + \frac{2a}{5}\right)
\phin ^6  (\tau)\right]\right\} . \label{m7}
\end{eqnarray}
To the same order in $g$,
the Jacobian action (\ref{m4}) is
\begin{eqnarray}
 &&\!\! {\cal A}_J[\phin ]=
-\delta (0)\int d \tau\left[-g \phin ^2(\tau) +
g^2 \left(a - \frac{1}{2}\right) \phin ^4 (\tau)
\right] . \label{m8}
\end{eqnarray}
For $g=0$,
the transformed partition function
(\ref{m2}) coincides with (\ref{mq2}).
When expanding   $Z$ of Eq.~(\ref{m2})
in powers of $g$,
we obtain
sums of Feynman diagrams contributing to each order $g^n$, which must vanish
to ensure coordinate invariance.
By considering only
connected Feynman diagrams,
we are dealing directly with the ground state energy.

\section{Free Energy Density}
The graphical expansion for
the ground state energy
will be carried here only up to three loops.
At
any order $g^n$, there exist different types
Feynman diagrams with $L=n+1, n,$ and $n-1$
number of loops
coming from the interaction
terms (\ref{m7}) and (\ref{m8}), respectively.
The diagrams are composed of the
three
types of lines in (\ref{p1})--(\ref{p3}), and
new interaction vertices
for each
power of $g$. The diagrams coming from the Jacobian action
 (\ref{m8}) are easily recognized by an accompanying power of $ \delta (0)$.
\comment{All divergences are parametrized by
Dirac delta function and its powers at origin,
so that the divergent piece of each graph
is easy to recognize.}

At first order in $g$, there
exists only three diagrams, two
originated from the
interaction (\ref{m7}), one
from the Jacobian action (\ref{m8}):
\begin{equation}
-\,g\hspace{0mm}\raisebox{-1mm}{\mbox{\input 6.tps }} - g\,\omega^{2}\hspace{-27mm}\raisebox{-11.57mm}{\mbox{\input inf.tps }} ~~~~~~~~~~~~~~
~~~
~~~
~~~
 + g\,\delta (0) \hspace{0mm}\raisebox{-1mm}{\mbox{\input 0dot.tps }} .
\label{f1}\end{equation}
~\\[-1.2cm]
At order $g^2$,
we distinguish several contributions.
First there are two three-loop local diagrams
coming
from the interaction (\ref{m7}),
and one two-loop local diagram
from the Jacobian action (\ref{m8}):
\begin{eqnarray}
 &&  g^2\,\Bigg[\,\,\,
 3 \left(\frac{1}{2} + a\right)\hspace{0mm}\raisebox{-3.2mm}{\mbox{\input 7.tps }} +
\,15 \omega^{2} \left(\frac{1}{18} + \frac{a}{5}\right)
 \hspace{-27mm}\raisebox{-13.7mm}{\mbox{\input clover.tps }}~~~~~~~~~~~~~~~~~~~~~~
~\\[-1.2cm] \nonumber\\
&&~~~ - 3\left(a - \frac{1}{2}\right)\, \delta (0)\,\,
\hspace{-27mm}\raisebox{-11.57mm}{\mbox{\input inf.tps }} ~~~~~~~~~~~~~~
~~~~~~~~~
\,\,\, \Bigg]\,.
\label{f2}\end{eqnarray}
~\\[-1.2cm]
We call a diagram local if it involves
no temporal time integral.
The Jacobian
action (\ref{m8}) contributes further
the nonlocal diagrams:
\begin{eqnarray}
 &&\!\!
-\frac{g^2}{2!}\bigg\{
 2\delta^2 (0) \!\!\hspace{0mm}\raisebox{-1mm}{\mbox{\input 0dotdot.tps }}
 \!\!- 4\delta (0) \big[\!\!
\hspace{0mm}\raisebox{-1mm}{\mbox{\input 6dot.tps }}\!+\!\!\!\!\!
\hspace{0mm}\raisebox{-1mm}{\mbox{\input 6pdot.tps }}
\!\!+ 2\,\omega^{4}\!\!\hspace{0mm}\raisebox{-1mm}{\mbox{\input infdot.tps }}\big]\bigg\}.
\nonumber\\
\label{f3}\end{eqnarray}
The remaining diagrams
come from the interaction  (\ref{m7}) only.
They  are either of the
three-bubble type, or of the watermelon
type, each with all possible combinations
of the three line types (\ref{p1})--(\ref{p3}):
The sum of all three-bubbles diagrams
is
\begin{eqnarray}
&& -\frac{g^2}{2!}\big[~~ 4\hspace{0mm}\raisebox{-1.2mm}{\mbox{\input 8.tps }}
\!+\,\,2~\hspace{0mm}\raisebox{-1.2mm}{\mbox{\input 9.tps }}
~+\,\,\,2\,\,\,\hspace{0mm}\raisebox{-1.2mm}{\mbox{\input 10.tps }}
\nonumber \\
&&~~~+
 ~8\,\omega^2\!
 \hspace{-27.0mm}\raisebox{-11.5mm}{\mbox{\input threeb1.tps }}~~~
~~~~~~~~~~~~~~~~~~~
+ 8\omega^2 \!\!\hspace{-27.0mm}\raisebox{-11.5mm}{\mbox{\input threeb2.tps }}~~~~~~~~~~~~~~~~~~~~~~~
 + 8 \omega^4 \hspace{-27.0mm}\raisebox{-11.5mm}{\mbox{\input threeb.tps }}~~~~~
\quad\quad\quad\quad\quad \quad \big] ~,
\label{f4}\end{eqnarray}
~\\[-1.2cm]
while the watermelon-like diagrams
contribute
\begin{eqnarray}
& &
{\!\!-\frac{g^2}{2!}\, 4 \,\bigg[\!\!
\!\!\!\hspace{0mm}\raisebox{-2mm}{\mbox{\input 11.tps }}
\!\!+ 4\!\!\!\hspace{0mm}\raisebox{-1.95mm}{\mbox{\input 12.tps }}
\!\!+\!\!\!\!\! \hspace{0mm}\raisebox{-1.9mm}{\mbox{\input 13.tps }}
\!\!+  4\omega^2\! \hspace{-27.0mm}\raisebox{-12.3mm}{\mbox{\input waterm2.tps }}
~~~~~~~~~~~
~~~~~~~~~~~
 \!+\! \!\frac{2}{3}\omega^4 \!
\hspace{-27.0mm}\raisebox{-12.3mm}{\mbox{\input waterm.tps }}
~~~~~~~~~~~
~~~~~~~~~~~
\bigg].   \!\!\!\!\!        }
\label{f5}
\end{eqnarray}
~\\[-01.1cm] Since the equal-time
expectation value $\langle\dot\phin (\tau)\,\phin (\tau)\rangle$
vanishes according to Eq.~(\ref{p5}),
there are a number of trivially vanishing diagrams,
 which have been omitted.

In our
previous papers \cite{2,3},
all
integrals were calculated individually
in
$D=1-\varepsilon$ dimensions,
taking the limit
$\varepsilon\rightarrow 0$ at the end.
The results for the integrals ensured that the sum of all
Feynman diagrams
contributing to each order $g^n$ vanishes.
\comment{Here we
derive the rules for
calculating partial sums
of Feynman diagrams directly
in one dimension from
the reparametrization
invariance of path integrals.}

\section{Rules for Integrals over Distributions}
As a first step in calculating the
Feynman integrals
we express
singular  time derivatives
$\dot\Delta (\tau)$, $\ddot\Delta (\tau)$
in terms
of regular correlation  functions
$\Delta (\tau)$, plus integrals over powers of $\delta $-functions.
The tools for this
will be
partial integrations and
the inhomogeneous field
equation satisfied by the correlation  function
\begin{equation}
\ddot \Delta  (\tau) = - \int \db k\, \frac{k^2}{k^2+ \omega^2}
 e^{ik\tau} = - \, \delta (\tau) + \omega^2  \Delta (\tau)\,,
\label{i1}\end{equation}

Most simply, we have for integrals over products
of two correlation  functions
the relation
\begin{equation}
\!\! \int d \tau\,\left[\dot\Delta^2 (\tau)
+ \omega^2  \Delta ^2 (\tau)\right] =
   \Delta (0).
\label{i2}\end{equation}
To prove this, we integrate
the first term partially,
\begin{equation}
 \int d \tau \, \dot\Delta^2 (\tau) =
 -  \int d \tau\, \Delta (\tau)\ddot\Delta (\tau),
\label{nolabel}\end{equation}
with no boundary term due to the exponential
vanishing at infinity of all functions involved.
Using now the field equation (\ref{i1})
and the
property
of $\delta$-function that
\begin{equation}
\!\! \int d \tau\,f (\tau)\,\delta (\tau) =
   f (0),
\label{nplabel}\end{equation}
for any smooth
test function $f (\tau)$,
we obtain (\ref{i2}).

We now turn to singular integrals
involving
$\ddot\Delta^2 (\tau)$.
Using the same tools, we
obtain, in the same way,
the relation
\begin{eqnarray}
&&
\int\! d \tau \!\left[\ddot\Delta^2 (\tau)
+ 2\omega^2 \dot \Delta ^2 (\tau)
+\omega^4  \Delta ^2 (\tau)
\right]
 \!=\! \int d \tau \,
\delta^2 (\tau)\,.
\label{i3}\end{eqnarray}

The last integral is undefined.
Before fixing its value in the next section,
we shall
derive relations
for integrals over singular products
of four correlation  functions.
First for $\ddot  \Delta (\tau ) \Delta ^3(\tau )$.
Using again the field equation (\ref{i1}), we find
\begin{equation}
\!\!- \int d \tau \, \ddot\Delta (\tau) \Delta ^3 (\tau) =
  \Delta ^3 (0) - \omega^2 \int d \tau\, \Delta ^4 (\tau).
\label{i4}\end{equation}
By a partial integration, the left-hand side becomes
\begin{equation}
 \int d \tau \,\ddot\Delta (\tau) \Delta ^3 (\tau)
 = - 3 \int d \tau \, \dot\Delta ^2 (\tau) \Delta ^2 (\tau) ,
\label{nolabel}\end{equation}
leading to
\begin{equation}
\int d \tau\, \dot\Delta ^2 (\tau) \Delta ^2 (\tau) =
\frac{1}{3}\Delta ^3 (0) -
\frac{1}{3}\omega^2 \int d \tau \,\Delta ^4 (\tau) .
\label{i5}\end{equation}
Invoking once more the field equation (\ref{i1}),
we obtain
the integral
\begin{equation}
\int d \tau \,\ddot\Delta (\tau) \dot\Delta ^2 (\tau) \Delta (\tau) =
 \omega^{2} \int d \tau\,\dot\Delta ^2 (\tau) \Delta ^2  (\tau) ,
\label{i7}\end{equation}
where we have used $\dot\Delta (0) = 0$.
Due to Eq.~(\ref{i5}), this
takes the form
\begin{equation}
\int d \tau\,\ddot\Delta (\tau) \dot\Delta^2 (\tau) \Delta (\tau) =
   \frac{1}{3} \omega^2  \Delta ^3 (0) - \frac{1}{3}
  \omega^4 \int d \tau\, \Delta ^4 (\tau) .
\label{i8}\end{equation}
A further partial integration
reduces
the integral
\begin{eqnarray}
\int d \tau\,\dot\Delta^4 (\tau)
= -3 \int d \tau \,\Delta (\tau ) \dot\Delta^2 (\tau)
 \ddot\Delta  (\tau)
\label{i9}
\end{eqnarray}
to (\ref{i8}), such that we arrive
at the relation
\begin{equation}
 \int d \tau\,\dot\Delta^4 (\tau) =
   - \omega^2  \Delta ^3 (0) +
  \omega^4 \int d \tau\, \Delta ^4 (\tau) .
\label{i10}\end{equation}

We now consider an integral over $\ddot  \Delta^2 (\tau ) \Delta ^2$.
Applying again the field equation (\ref{i1}),
we find    the relation
\begin{eqnarray}
\int d \tau\,\ddot\Delta ^2 (\tau) \Delta ^2 (\tau)& = &
 \int d \tau\,
\Delta^2 (\tau)\delta^2 (\tau)
\nonumber\\& - &
2 \omega^2  \Delta^{3} (0)
  + \omega^4 \int d \tau \, \Delta ^4 (\tau),
\label{i6}\end{eqnarray}

The relations Eqs.~(\ref{i3}) and (\ref{i6})
have
reduced all
integrals over singular
products of  correlation  functions
to regular integrals plus two undefined integrals containing
$ \delta ^2(\tau )$.
We are now going to show, that
the reparametrization
invariance of path integrals
requires the following rules
for integrals over
products of two $ \delta $-functions in Eqs.~(\ref{i3}) and (\ref{i6}):
\begin{equation}
\int d \tau\,\delta^2 (\tau) =  \delta (0) ,
\label{i11a}\end{equation}
and  further
\begin{equation}
\int d \tau\, f (\tau)\delta^2 (\tau) = f (0)\delta (0),
\label{i11b}\end{equation}
for any smooth test function  $f(\tau )$.

\section{Imposing Reparametrization Invariance}
To first order in $g$,
the sum of
Feynman diagrams (\ref{f1})
must vanish:
\begin{equation}
\hspace{0mm}\raisebox{-1mm}{\mbox{\input 6.tps }} + \,\omega^{2}\hspace{-27mm}\raisebox{-11.57mm}{\mbox{\input inf.tps }} ~~~~~~~~~~~~~~
~~~
~~~
~~~
 -\,\delta (0) \hspace{0mm}\raisebox{-1mm}{\mbox{\input 0dot.tps }} = 0 .
\label{di1}\end{equation}
~\\[-1.2cm]
The analytic form of this relation is
\begin{equation}
\left[
- \ddot \Delta  (0) + \omega^2  \Delta (0)
  - \, \delta (0)\right]\,\Delta (0) = 0 ,
\label{ai1}\end{equation}
and the vanishing is a direct consequence
of the field
equation  (\ref{i1})
 for the correlation  function
at origin.

At order $g^2$,
the same equation
reduces the sum of all local diagrams in
(\ref{f2}) to a finite result plus a
term proportional to $ \delta (0)$:
\begin{eqnarray}
&& \left[- 3 \left(\frac{1}{2} + a \right) \ddot\Delta (0) +
15 \left(\frac{1}{18} + \frac{a}{5}\right)
\omega^2 \Delta (0)\right.
\\ && \left.
- 3 \left(a - \frac{1}{2}\right)\delta (0)
\right]\Delta^2 (0)
 = \,\left[3 \delta (0)
 - \frac{2}{3} \,\omega^2 \Delta (0)\right]\Delta^2 (0).
\nonumber
\label{ai4}\end{eqnarray}
Representing  right-hand side diagrammatically, we obtain
the identity
\begin{equation}
\!\!\!\!\mathop{\Sigma }{(\ref{f2})} =
  3 \delta (0)\hspace{-27mm}\raisebox{-11.57mm}{\mbox{\input inf.tps }} ~~~~~~~~~~~~~~
~~~~~~~~~
- \,\frac{2}{3}\, \omega^{2}
 \hspace{-27mm}\raisebox{-13.7mm}{\mbox{\input clover.tps }}~~~~~~~~~~~~~~~~~~~~~~ ,
\label{di4}\end{equation}
~\\[-1cm]
where
$\mathop{\Sigma }{(\ref{f2})}$ denotes the sum of all diagrams in Eq.~(\ref{f2}).
 Using the identity (\ref{i2})
together with the field equation (\ref{i1}),
we reduce the sum (\ref{f3})
of all one and two-loop bubbles
diagrams to terms involving $ \delta (0)$ and $ \delta ^2(0)$:
\begin{eqnarray}
&& -  \frac{1}{2!}
\left\{ 2\delta^2 (0) \int \!\!d \tau\,\Delta^2 (\tau)
\right.\nonumber\\
&  &\left. -4\delta (0)\!\! \int \!\!d \tau\,\left[
\Delta (0)\dot\Delta^2 (\tau)
\!-\! \ddot\Delta (0) \Delta^2 (\tau)
 \!+\! 2\omega^2 \Delta (0)\Delta^2 (\tau)\right]\right\}
 \nonumber\\
& &=  2\Delta^2 (0)\,\delta (0)
 +  \delta^2 (0)\int d \tau \Delta^2 (\tau) .
\label{ai2}\end{eqnarray}
Hence we find  the diagrammatic identity
\begin{equation}
 -\frac{1}{2!}\,\mathop{\Sigma }{(\ref{f3})} =
 2\delta (0)\hspace{-27mm}\raisebox{-11.57mm}{\mbox{\input inf.tps }} ~~~~~~~~~~~~~~
~~~~~~~~~
+\,\, \delta^2 (0) \!\!\hspace{0mm}\raisebox{-1mm}{\mbox{\input 0dotdot.tps }} .
\label{di2}\end{equation}
~\\[-1.2cm]
Now, the terms accompanying $ \delta ^2(0)$
turn out
to be canceled by
similar terms coming from the sum
of all three-loop bubbles
diagrams in (\ref{f4}).
In fact, the
identities (\ref{i2}) and (\ref{i3}) lead to
\begin{eqnarray}
&&
- \frac{1}{2!}\int d \tau\,\left[
-4\Delta (0)\ddot\Delta (0)\dot\Delta^2 (\tau)
+2\Delta^2 (0)\ddot\Delta^2 (\tau)
\right. \nonumber\\
&&~~~~~~+2\ddot\Delta^2 (0)\Delta^2 (\tau)
+ 8\omega^2 \Delta^2 (0)\dot\Delta ^2 (\tau)\nonumber\\
&  &~~~~~~- \left. 8\omega^2 \Delta (0)\ddot\Delta (0)\Delta^2 (\tau)
 +8\omega^4 \Delta^2 (0) \Delta^2 (\tau)
\right]
\\
&&
~~= ~-\left[\int d \tau\,\delta^2 (\tau) +  2 \delta (0)\right]\, \Delta^2 (0)
 - \delta^2 (0)\,\int d \tau \, \Delta^2 (\tau) . \nonumber
\label{ai3}\end{eqnarray}
Thus, we find the diagrammatic
identity for all bubbles diagrams
\begin{equation}
 -\frac{1}{2!}\,\mathop{\Sigma }{(\ref{f3})}
 -\frac{1}{2!}\,\Sigma(\ref{f4}) =
 - \int d \tau\,\delta^2 (\tau)\hspace{-27mm}\raisebox{-11.57mm}{\mbox{\input inf.tps }} ~~~~~~~~~~~~~~
~~~~~~~~~\,.
\label{di3}\end{equation}
~\\[-1.2cm]
Finally, the relations  (\ref{i5}), (\ref{i6}),
(\ref{i8}) and (\ref{i10}) reduce
the sum (\ref{f5})  of all watermelon-like diagrams
to a finite contribution
plus the integral involving  $\delta^2 (\tau)$:
\begin{eqnarray}
&  &\!\!\!\!\!\!- \frac{4}{2!}\int d \tau\,\left[
\Delta^2 (\tau) \ddot\Delta^2 (\tau)
+4\Delta (\tau) \dot\Delta^2 (\tau) \ddot\Delta (\tau)\phantom{\int}
\right. \nonumber\\
&&~~~~~~~~+ \left.
\dot\Delta^4 (\tau)
 + 4\omega^2 \Delta^2 (\tau) \dot\Delta ^2 (\tau)
+ \frac{2}{3}\omega^4 \Delta^4 (\tau)
\right] \nonumber \\
&&~~~~= ~- ~2 \int d \tau\,\Delta^2 (\tau) \delta^2 (\tau)
 + \frac{2}{3}\,\omega^2 \Delta^3 (0)\,.
\label{ai5}\end{eqnarray}
Combining these with all local
diagrams (\ref{di4}), we
easily verify
that all finite contributions
cancel each other leading to
the diagrammatic identity
\begin{eqnarray}
&&\!\!\!\!\!\!\mathop{\Sigma }{(\ref{f2})}
 -\frac{4}{2!}\Sigma(\ref{f5})
 \nonumber\\&&
 =\left[ 3\delta (0)
 - 2 \Delta^{-2} (0)
 \int d \tau\,\Delta^2 (\tau) \delta^2 (\tau)
 \right]
 \hspace{-27mm}\raisebox{-11.57mm}{\mbox{\input inf.tps }} ~~~~~~~~~~~~~~
~~~~~~~~~ \, .
\label{di5}\end{eqnarray}
~\\[-1.2cm]
If the
singular terms
in  Eqs.~(\ref{di3}) and (\ref{di5})
are to sum up to zero, as required by
the coordinate invariance of perturbatively
defined path integrals, we must
have the integration rules
for the square distribution
(\ref{i11a}) and (\ref{i11b}), which
determined completely
the right-hand sides of relations (\ref{i3})
and  (\ref{i6}).

The procedure can easily be continued to  higher-loop
diagrams to obtain integrals over any desired products of
singular correlation functions, and over products of  $ \delta $-functions.

At no place do we have to specify
the value of $ \delta (0)$ and the regularization scheme.
There is a  perfect cancellation
of all powers of $ \delta (0)$
arising from the expansion  of the Jacobian action,
and this
is the reason why the so-called
Veltman rule  of setting
 $\delta (0)=0$  can be used everywhere without problems.

\section{Summary}
In this note we have set up
simple rules for
relating singular to regular Feynman
integrals
which avoid the  explicit calculation of
dimensionally regularized integrals
over products of distributions.
These rules follow directly
from the invariance of
perturbatively defined  path integral
under coordinate transformations.
Our procedure
is {\em independent\/} of
of regularization
prescriptions, using only
the fact that regularized integrals can be integrated by parts.
The results are, of course,
perfectly compatible with those derived
before in Refs.~\cite{2,3} by
dimensional regularization.

Just as in the time-sliced definition of
path integrals
in curved spcae in Ref.~\cite{1},
there is absolutely no need for extra compensating potential terms
found necessary in the treatments in Refs.~\cite{5,7,6}.

\end{document}